# INFRARED ABSORPTION INVESTIGATIONS CONFIRM THE EXTRATERRESTRIAL ORIGIN OF CARBONADO-DIAMONDS.

Jozsef Garai[1,*], Stephen E. Haggerty[1],

[1] Dept. of Earth Sciences, Florida International University, PC-344, Miami, FL 33199, USA

Sandeep Rekhi[2,3,4], and Mark Chance[2,3]

[2] Case Western Reserve University, Cleveland, OH 44106-4988, USA

[3] Case Center for Synchrotron Biosciences, National Synchrotron Light Source, Brookhaven National Laboratory, Upton, New York 11973, USA

[4] Physics Department, Harvard University, Cambridge MA 02139

[*]Corresponding author. Tel: 786-247-5414; e-mail: jozsef.garai@fiu.edu

## ABSTRACT

The first complete infrared FTIR absorption spectra for carbonado-diamond confirm the interstellar origin for the most enigmatic diamonds known as carbonado. All previous attempts failed to measure the absorption of carbonado-diamond in the most important IR-range of 1000-1300 $cm^{-1}$ (10.00-7.69 μm) because of silica inclusions. In our investigation, KBr pellets were made from crushed silica-free carbonado-diamond and thin sections were also prepared. The 100 to 1000 times brighter synchrotron infrared radiation permits a greater spatial resolution. Inclusions and pore spaces were avoided and/or sources of chemical contamination were removed. The FTIR spectra of carbonado-diamond mostly depict the presence of single nitrogen impurities, and hydrogen. The lack of identifiable nitrogen aggregates in the infrared spectra, the presence of features related to hydrocarbon stretch bonds, and the resemblance of the spectra to CVD and presolar diamonds indicate that carbonado-diamonds formed in a hydrogen-rich interstellar environment. This is consistent with carbonado-diamond being sintered and porous, with extremely reduced metals, metal alloys, carbides and nitrides, light carbon isotopes, surfaces with glassy melt-like patinas, deformation lamellae, and a complete absence of primary, terrestrial mineral inclusions. The 2.6 – 3.8 billion year old fragmented body was of asteroidal proportions.

1. INTRODUCTION

Carbon is the most abundant dust-forming element in the interstellar medium. The dominant fraction of carbon is in the form of grains, comprised principally of hydrocarbons. Other components include graphite, hydrogenated amorphous aliphatic and/or aromatic hydrocarbons (a-C, a-C:H) and diamond. The existence of interstellar diamond was predicted by Saslaw & Gaustad (1969), and subsequently found in primitive, carbonaceous chondritic meteorites (Lewis et al. 1987). Interstellar diamonds are cubic, and coexist with highly reduced carbides of Si, Mo, W, and Ti. Although there is widespread consensus that diamonds can form in interstellar environments (Kouchi et al. 2005, Binette et al. 2005), the processes are debated (Lewis et al. 1987; Tielens et al. 1987, Daulton, Ozima, & Shukolyukov 1995, Nuth & Allen 1992, Duley & Grishko 2001).

The presence of diamond-related bands at 2915 and 2832 $cm^{-1}$ (3.43, and 3.53 μm) in the IR emission spectra of two Herbig Ae/Be stars HD 97048 and Elias 1 (Guillois, Ledoux, & Reynaud 1999; Van Kerckhoven, Tielens, & Waelkens 2002) are well established. A tentative peak at 2832 $cm^{-1}$ (3.53 μm) has also been detected in V921 Sco (=CD-42.11721), HD 163296 and T CrA (Ache, & Van den Ancker 2006). The bands are interpreted as surface C-H stretching modes of hydrogenated nanodiamonds, with particle-sizes in HD 97048 and Elias 1 of at least 50 nm (Jones et al. 2004).

Terrestrial diamonds are recovered from exotic volcanoes (kimberlites and lamproites) that originate at depths >150 km. The single exception is carbonado, a tough, black to dark grey polycrystalline cuboid diamond (Fig. 1), 2.6 – 3.8 Ga in age, found in the same sedimentary geological horizon in Brazil and the Central African Republic (Haggerty, 1998; Heaney, Vicenzi, & De 2005 and refs. therein). Surfaces are commonly patinated and samples are coke or ceramic-like (Haggerty, 1998) with vesicles of up to 1 mm (Fig. 1) and porosities of up to ~30% (Vicenzi et al. 2003); the C isotopic composition ($\delta^{13}C$‰ is between -21 and -34) is extremely light (Russell, Arden, & Pillinger 1996; Shelkov et al. 1997); planar-defect shock lamellae are present (Heaney, Vicenzi, & De 2005); and the diamonds contain an extraordinary array of highly reduced metals and metal alloys (Fe, Ni, Fe-Ni, Fe-Cr, Ni-Cr), SiC, and TiN (De and Heaney, 1996; Gorschkov et al. 1996; Jones, Milledge, & Beard 2003; Parthasaratha et al. 2004). Because none of the above characteristics are compatible with a deep Earth environment, a reasoned alternative is that carbonado-diamond is of extraterrestrial origin (Haggerty, 1996, 1998, 1999). This proposal is furthered explored here in the context of the IR spectra of carbonado.

An important physical characteristic of diamond, which may shed light on its origin, is the aggregation level, as outlined in section 3.2, of substituted nitrogen (Table 1). The IR absorption in the most important nitrogen related region, between 1000-1300 cm$^{-1}$ (10.00-7.69 μm), has never been collected from carbonado-diamond. The pores of carbonado are typically stuffed with silica-based minerals that yield a broad absorption feature between 900 and 1300 cm$^{-1}$ (11.11 and 7.69 μm). The intensity of this feature is much stronger than the defect-activated one phonon bands, thus preventing the measurement of the infrared absorption of carbonado-diamond (Kagi et al. 1994; Magee & Taylor 1999). No experiments have succeeded in overcoming these problems and the diamond type (e.g. Ia, IaA, IaB, II) in carbonado has yet to be determined.

Electron paramagnetic resonance of carbonado-diamond shows the presence of single substitutional nitrogen centers, labeled P1 (Smith et al. 1959; Braatz et al. 2000; Nadolinny et al. 2003). Defects containing hydrogen, labeled H1, and higher N aggregation centers, $N_2^+$, $N_2V^+$, and $N_3V$, reported by Nadolinny, et al. (2003), are, however, not observed by others (e.g. Braatz et al. (2000).

## 2. EXPERIMENTAL

In order to remove silica contamination from the carbonado-diamonds, two Brazilian and two Central African samples (3-5 gm) were crushed in tungsten carbide piston cylinder crusher to < 67μm. The resulting powders were treated with HCl for three days and then placed in HF pressure bombs at 60 $^{o}$C for an additional three days. The thoroughly washed, recovered carbonado-diamond was mixed with KBr in different proportions and then pressed under low vacuum to form ~1 mm thick pellets. A control pellet, with the same weight proportion was made from natural diamond crushed to ~25 μm. Three thin sections of carbonado-diamond, two from the Central African Republic and one from Brazil, were also prepared with no chemical treatment.

The synchrotron IR experiment was carried out at the U2b beam line located at the National Synchrotron Light Source at Brookhaven National Laboratory (NSLS-BNL). The beam line is equipped with a dry nitrogen purged FTIR spectrometer (Nicolet Magna 860) with KBr beamsplitter and a narrow band mercury cadmium telluride (MCT/A) detector coupled with an IR microscope (Nic-Plan, Nicolet Instruments, Madison, WI) with 32X IR objective (Thermo Spectra-Tech) and diffraction limited resolution. The spectrometer is equipped with a Michelson interferometer having capabilities such as rapid scan, step scan and the ability to accept external beams and detector. The bench is configured to use a collimated synchrotron light (beam energy 800 MeV) through an external input of the spectrometer. The IR spectra on various carbonado samples mixed with dry KBr were collected with a beam spot diameter around 100 μm. The background spectra (1064 scans) were collected and subtracted from the IR spectra of samples. The IR spectra were collected in the mid-IR range 4000–650 cm$^{-1}$ (2.50-15.34 μm) at a resolution of 4 cm$^{-1}$ with 1056 scans co-added. Stage control, data collection and processing were performed using OMNIC 7.2 (Thermo-Nicolet, Madison, Wisconsin). The portions of carbonado samples were scanned and selected using a CCD camera linked to the infrared images (objective 32X).

## 3. RESULTS AND DISCUSSION

Using a glowbar (black body) generated source, the infrared absorption spectra of the pellets were collected from 650-4000 cm$^{-1}$ (15.34-2.50 μm). Outside the nitrogen absorption region 1000-1500 cm$^{-1}$ additional bands in the 650-1000 cm$^{-1}$ (15.34-6.67 μm) region, and a triplet around 2900 cm$^{-1}$ (3.45 μm) were observed in the carbonado-diamond spectra. The additional non - nitrogen absorption peaks are interpreted as C-H related bands.

Using a synchrotron radiation source with a 100 times better signal/noise ratio than the laboratory instrument using a glowbar infrared source (Reffner et al. 1994) the absorption spectra of the thin sections were collected from 650-4000 cm$^{-1}$ (15.34-2.50 μm). The prominent peaks present in all spectra are 1102, 2855, 2926, and 2961 cm$^{-1}$ (9.07, 3.50, 3.42, and 3.38 μm) (Fig. 2; Table 2). Additional sharp peaks are also present but are not developed in all runs. There are relatively broad features, present in the three thin section investigations, at 1660, 2030, 2155 cm$^{-1}$ (6.02, 4.93, 4.64 μm).

Any possibility of contamination from the instrument and/or contamination by organic solvents was ruled out by taking the spectra of natural diamond, which did not show C-H related peaks.

### 3.1. *C-H Stretching Bands*

The peak frequencies (Fig. 2) at 2855, 2926, and 2961 cm$^{-1}$ (3.50, 3.42, and 3.38 μm) can be attributed to C-H stretching of diamond hydride and are sensitive to the surface structure (Chin et al. 1995; Alfonso, Drabold, & Ulloa 1995; Zhigilei, Srivastava, & Garrison 1997). Hydrogen ionization experiments on diamond single crystals show a band at 2835 cm$^{-1}$ (3.53 μm) that is ascribed to C(111)-1 x 1 : H, (i.e. single carbon – hydrogen bonds on octahedral (111) crystallographic faces), a doublet at 2921 and 2935 cm$^{-1}$ (3.42, and 3.41 μm), associated with C-H on (100), and 2857 and 2874 cm$^{-1}$ (3.50, and 3.48 μm) bands related to (110) (Chin et al. 1995; Cheng et al. 1996). The complete assignment of C-H stretching in polycrystalline diamonds, however, is incomplete.

The (111) surface peak is absent in our spectra. The prominent bands at 2855, and 2926 cm$^{-1}$ (3.50, and 3.42 μm), are assigned respectively to CH$_x$ and C{100} (Guillois, Ledoux, & Reynaud 1999; Jones et al. 2004). The band observed at 2961 cm$^{-1}$ (3.38 μm), and reported in other studies, remains unassigned. The lack of the C(111)-1 x 1 : H surface, but the presence of the C{100} surface in the carbonado-diamond spectra is fully consistent with the prevalent cuboid habit of individual diamond crystals in our carbonado collection. And it is consistent with *asteriated* (zoned and star-shaped) diamonds where cuboidal forms have a strong preference for hydrogen, and octahedral segments for nitrogen (Rondeau et al. 2004). More generally, hydrogen is critical to diamond formation in CVD processes, is present as *"dangling bonds"* (i.e. singly occupied outwardly-directed sp$^3$ hybrid orbitals associated with each surface carbon atom) in natural diamonds (Evans, 1992), is invoked in *"hydrogenated nano-diamonds"* in meteorites (Andersen et al. 1998), and with *"protonation"* as the bonding agent that sinters microdiamonds in carbonado (Haggerty, 1998).

### 3.2. *Nitrogen Related Bands*

Nitrogen (Z=7) may replace carbon (Z=6), but long term residence at high temperatures results in N migration and high aggregation levels (Evans, & Qi 1982); short time residence at low temperatures, or quenched CVD diamonds, correspond primarily to single N as a substitutional impurity. We observe a prominent absorption peak (present in all of the spectra) in the C-N stretching region at 1102 cm$^{-1}$ (9.07 μm), which is also reported in the IR spectra of presolar diamonds from the Allende meteorite (Lewis et al. 1989) and in CVD diamonds (Andersen 1998). The peak is attributed to C-N stretching and/or C-O stretching in aliphatic ethers (Koike et al. 1995; Colangeli et al. 1994; Mutschke et al. 1995; Hill et al. 1997).

The peak at 1102 cm$^{-1}$ (9.07 μm) is relatively wide (FWHM = 103.6 cm$^{-1}$) and asymmetric with a shoulder, indicative of two peaks. Spectral analyses show peaks at 1100 cm$^{-1}$ (9.09 μm) and 1128 cm$^{-1}$ (8.86 μm) which reproduce the observed spectra. The 1128 cm$^{-1}$ is attributed to substitutional nitrogen. The presence of a single substitutional nitrogen [C-N]$^o$, or P1 center in the carbonado-diamond is well established from electron paramagnetic resonance investigations (Smith et al. 1959; Braatz et al. 2000; Nadolinny et al. 2003) adding additional support to the existence of this peak. Hydrogen may have an effect on peak shifts and the influence of deformation or radiation has not been fully explored.

In the spectra of CAR-24, which has the most intensive 1102 cm$^{-1}$ peak, the accompanying asymmetric C-O-C stretch peak at 1223 cm$^{-1}$ (8.18 μm) is present, supporting the C-O-C origin for the peak. On the other hand, the possible nitrogen attribution of this band is supported by the increasing intensity of the band with nitrogen doping (Titus et al. 2006).

The other substitutional nitrogen related peak at 1350 cm$^{-1}$ (7.41 μm) is not present in our spectra, which is consistent with its absence in the reported spectra of presolar and CVD diamonds (Anderson et al. 1998). Another, possible nitrogen-related peak at 1285 cm$^{-1}$ (7.78 μm) is observed only in the spectra of CAR-6; its position, if neither red nor blue shifted, could be ascribed to A aggregation of N pairs. Two N atoms and a vacancy (designated as 3H and H3 centers), are typically the result of radiation or deformation, and are present in the photoluminescence spectra of carbonado-diamond (Clark, Collins, & Woods 1992; Nadolinny et al. 2003). There is no indication for higher order nitrogen centers or for N-platelets.

The absorption peak at 1384 cm$^{-1}$ (7.22 μm), reported by Kagi et al. (1994), has not been confirmed by other investigators. The position of the platelet absorption peak correlates with nitrogen abundance and inversely with platelet size; therefore, the 1384 cm$^{-1}$ (7.22 μm) absorption peak should be due to small platelets sizes and with abundant nitrogen. Carbonado-diamond, however, has low nitrogen contents (100-300 ppm) that should result in large platelets and a corresponding red shift rather than the blue shift suggested by Kagi et al. (1994). The absorption peak of aggregated platelets (at ~1370 cm$^{-1}$ ~7.30 μm) is always accompanied by an A absorption peak at 1282 cm$^{-1}$ (7.80 μm) (Davis 1977; Mendelssohn & Milledge 1995), but is absent in the spectra collected by Kagi et al. (1994). The absence of the A aggregates (pairs) in the spectra raise the possibility that the 1384 cm$^{-1}$ absorption peak might not originate from diamond. One explanation might be that the 1384 cm$^{-1}$ peak could be due to BN contamination (Mendelssohn & Milledge 1995) in the stainless container used to crush the carbonado (Kagi et al. 1994).

### 3.3. *Other Peaks*

The broad band features at 2030 and 2155 cm$^{-1}$ (4.93, 4.64 μm) are most likely related to diamond double phonon absorptions (Klein, Hartnett, & Robinson 1992). The 1660 cm$^{-1}$ (6.02 μm) band can be assigned to C=C. The remaining bands with well developed peaks from 800-1500 cm$^{-1}$ (12.5-6.67 μm) can be assigned to PAH-polycyclic aromatic hydrocarbons (Allamandola, Tielens, & Barker 1989). The broad feature in CAR-6 at 3195 cm$^{-1}$ (3.13 μm) can be attributed to N-H stretching.

The presence of PAH bands in the spectra is consistent with carbonado-diamonds containing ~0.002-0.004 wt. % (Kaminsky 1991); hydrocarbon peaks were also obtained from the thin sections, confirming that these are not an artifact of hydrofluoric acid treatment.

## 4. CONCLUSIONS

From the first complete infrared FTIR absorption spectra for carbonado-diamond we show that hydrogen stretching bands at 2855, 2926, and 2961 cm$^{-1}$ (3.50, 3.42, and 3.38 μm) are present. The strongest absorption band at 1102 cm$^{-1}$ (9.04 μm) is most likely due to substitutional nitrogen in hydrogenated diamond; however, the possibility of a C-O-C vibration cannot be excluded. All of the prominent peaks are present in treated and untreated samples and are intrinsic to carbonado-diamond. These bands may be useful in the identification of extraterrestrial diamond.

The identified absorption peaks, with the exception of some PAH related bands, are almost identical with the spectra reported for presolar and CVD diamonds (Andersen et al. 1998; Braatz et al. 2000), indicating that these diamonds were most likely formed in similar, hydrogen-rich environments.

Based on the overwhelming evidence for a close resemblance of the FTIR spectra of carbonado-diamond to presolar diamonds in meteorites, and to interstellar diamond dust; the overlapping range of C isotopic compositions with meteoritic nanodiamonds; the high porosity; the high concentration of polycyclic aromatic hydrocarbons; the cuboid shape micro-crystals; exotic metals, metal alloys, SiC, and TiN inclusions (found only in meteorites); and the high concentration of planar defect lamellae; the complete absence of a deep Earth mantle fingerprint, we conclude that the new IR measurements are consistent with an origin for carbonado-diamonds in an interstellar environment. The implication is that carbonado-diamond is of asteroidal proportions. This is entirely reasonably given that crystalline white dwarfs are interpreted as modified diamond (Metcalf, Montgomery, & Kanaan 2004), and that extrasolar carbon planets are posited to contain an inner, concentric horizon of diamond (Kuchner and Seager 2005).

## 5. ACKNOWLEDGMENTS

Support by the NSF (EAR95-05770), the Office of Grants and Contracts at the University of Massachusetts, and DeBeers for field work and laboratory studies to SEH is greatly appreciated. Thanks to Dr. Jen Bohon for helping in the collection of some IR spectra, and to the reviewers for suggested improvements.

TABLE I. COMMONLY OBSERVED BANDS IN THE IR SPECTRUM OF DIAMOND (McNamara 2003)

| Frequency (cm$^{-1}$) | Functional Group | Source/Reference |
|---|---|---|
| 1130 (8.85 μm) | substitutional N | Smith et al. 1959 |
| 1175 (8.51 μm) | B aggregate | Collins 1980 |
| 1220 (8.20 μm) | C-N stretch | Evans & Qui 1982 |
| 1250 (8.00 μm) | C-N, N-V | Davies, Lawson, & Collins 1992 |
| 1282 (7.80 μm) | A aggregates (pairs) | Kaiser & Bond 1959 |
| 1332 (7.51 μm) | Raman | intrinsic |
| 1350 (7.41 μm) | Substitutional N | Davies, Lawson, & Collins 1992 |
| ~1370 (~7.30 μm) | platelet N | Evans & Qui 1982; Evans, Qui, & Maguire 1981 |
| 1405 (7.12 μm) | overtone | Davis, Collins, & Spear 1984 |

TABLE II. THE ASSIGNMENT OF THE PROMINENT BANDS OBSERVED IN THE CARBONADO-DIAMOND FTIR SPECTRA.

| Frequency (cm$^{-1}$) | Assignment | Intensity | FWHM (μm) |
|---|---|---|---|
| 1102 ( μm) | C-N/C-O stretch | 1.00 | 0.347 |
| 2855 (3.50 μm) | CH$_x$ | 0.08 | 0.027 |
| 2926 (3.42 μm) | C{100} | 0.18 | 0.029 |
| 2961 (3.38 μm) | | 0.11 | 0.038 |

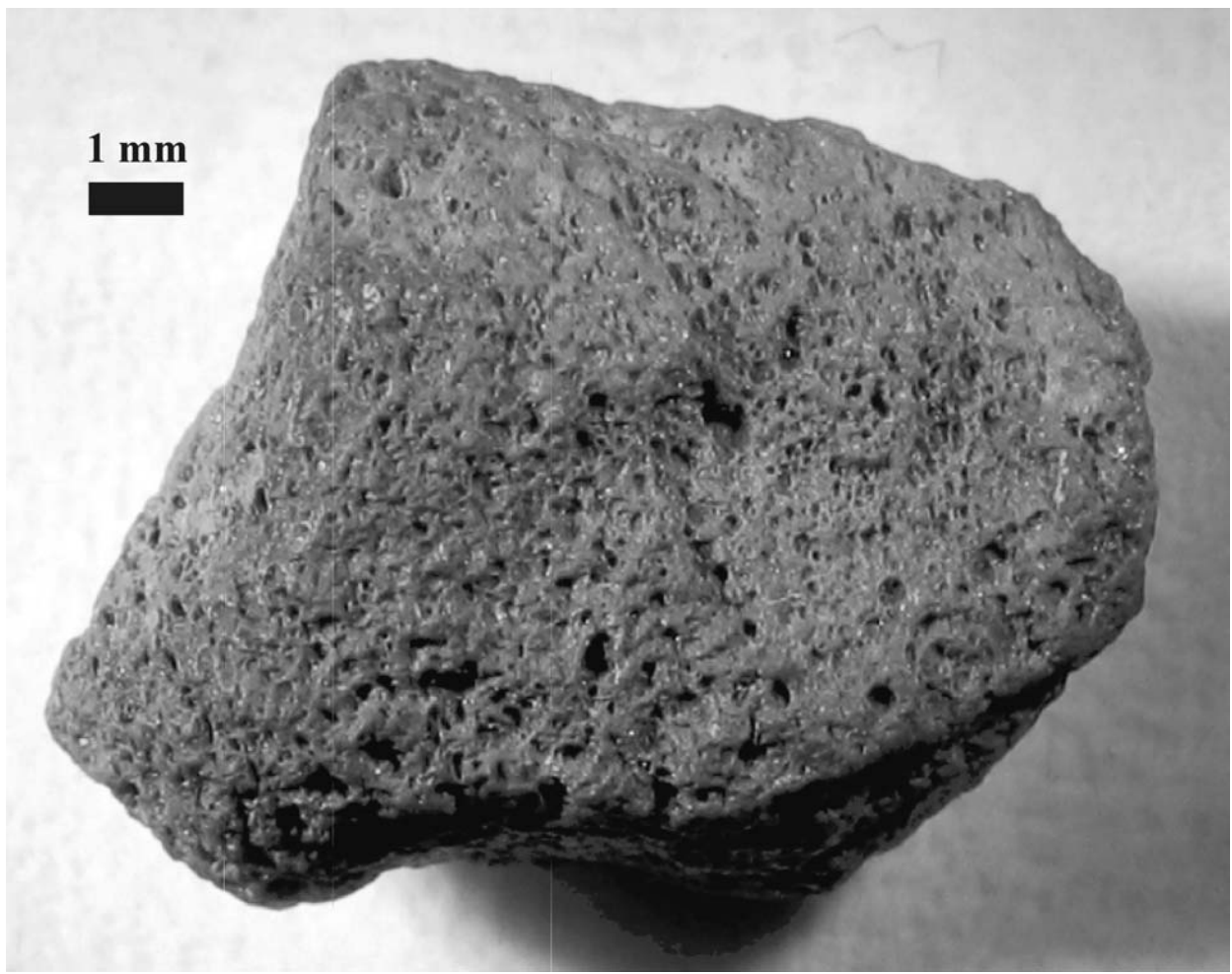

FIG. 1. A typical black and highly porous polycrystalline carbonado-diamond (5.3 cts) from Lencois, State of Bahia, Brazil. (Sample B-13).

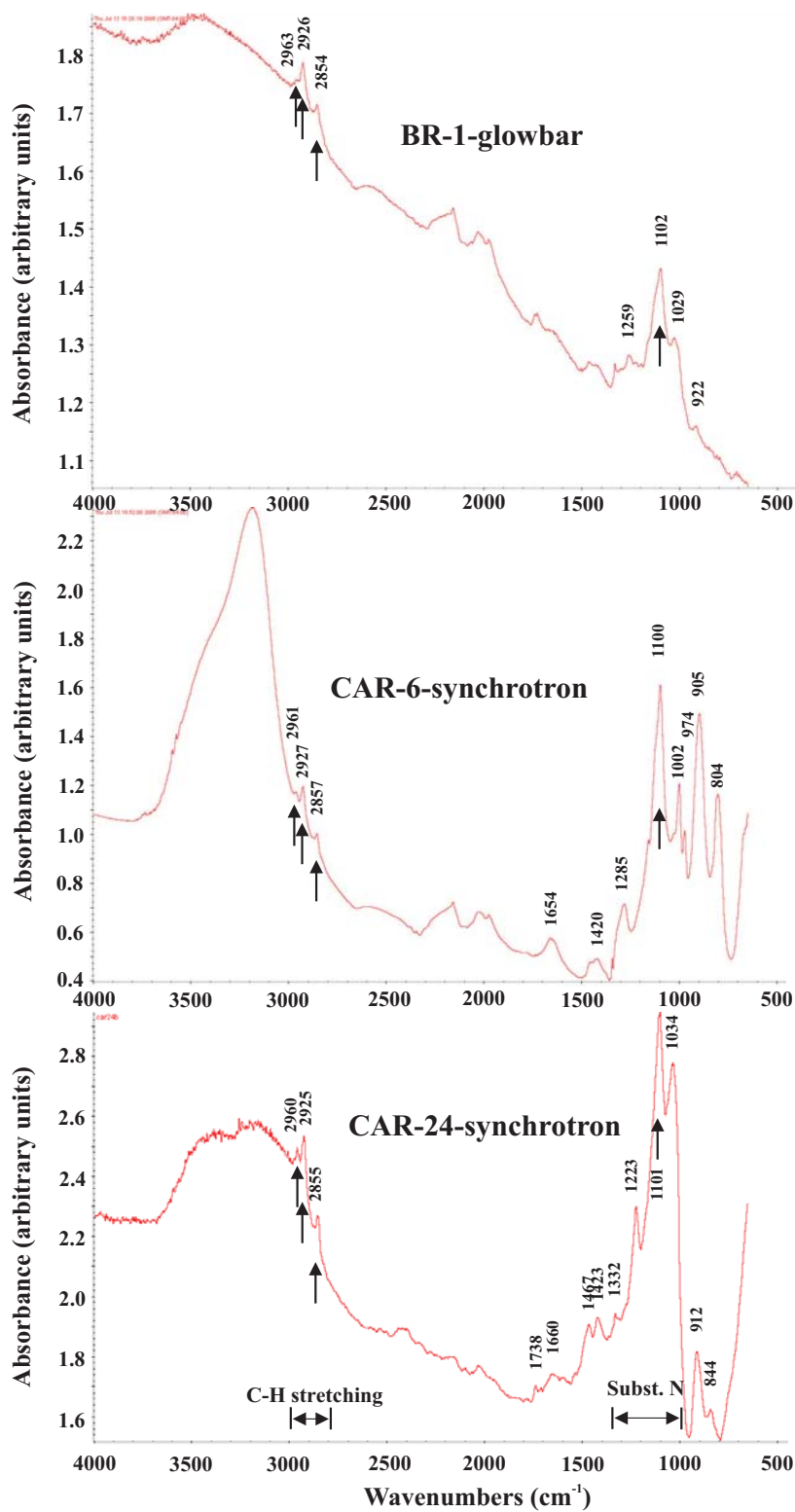

FIG. 2. Mid-infrared absorption spectra of carbonado-diamonds collected from a Central African (CAR) and Brazilian (BR) thin sections. The prominent peaks are marked with arrows and the assigned vibrational regions are identified.